**Title Page**

Major Classification: Physical Sciences

Title: Energy Conversion via Metal Nanolayers

Author Affiliation:

Mavis D. Boamah[a], Emilie H. Lozier[a], Jeongmin Kim[b], Paul E. Ohno[a], Catherine E. Walker[a], Thomas F. Miller III[b], and Franz M. Geiger[a,1]

[a]Department of Chemistry, Northwestern University, Evanston, IL 60208, USA and

[b]Department of Chemistry, California Institute of Technology, Pasadena, CA, 91125, USA

To whom correspondence may be addressed. Email: geigerf@chem.northwestern.edu

**Abstract.** Current approaches for electric power generation from nanoscale conducting or semi-conducting layers in contact with moving aqueous droplets are promising as they show efficiencies of around 30 percent, yet, even the most successful ones pose challenges regarding fabrication and scaling. Here, we report stable, all-inorganic single-element structures synthesized in a single step that generate electrical current when alternating salinity gradients flow along its surface in a liquid flow cell. 10 nm to 30 nm thin nanolayers of iron, vanadium, or nickel produce several tens of mV and several microA cm$^{-2}$ at aqueous flow velocities of just a few cm s$^{-1}$. The principle of operation is strongly sensitive to charge-carrier motion in the thermal oxide nano-overlayer that forms spontaneously in air and then self terminates. Indeed, experiments suggest a role for intra-oxide electron transfer for Fe, V, and Ni nanolayers, as their thermal oxides contain several metal oxidation states, whereas controls using Al or Cr nanolayers, which self-terminate with oxides that are redox inactive under the experimental conditions, exhibit dramatically diminished performance. The nanolayers are shown to generate electrical current in various modes of application with moving liquids, including sliding liquid droplets, salinity gradients in a flowing liquid, and in the oscillatory motion of a liquid without a salinity gradient.

209/250 words.

**Significance Statement**

This work reports kinetic:electrical energy transduction using nanolayers formed in a single-step from earth-abundant elements. The method utilizes large-area physical vapor deposition onto rigid or flexible substrates that can be readily scaled to arbitrarily large areas. In addition to flowing aqueous droplets across the nanolayers, current is shown to be created either with linear flow of salinity gradients or with oscillatory flow of a constant salinity. The operational requirement of having to move a dynamically changing electrical double layer (a "gate") across the nanostructure identified in prior approaches is confirmed for the new structures and augmented by a need for electron transfer within the thermal oxide nano-overlayers terminating the metals. The simplicity of the approach allows for rapid implementation.

119/120 words.

\body

**Main Text.** Previous methods for kinetic/gravitational to electrical energy conversion have used conducting or semi-conducting layered materials in contact with moving aqueous droplets or brushes. The most successful approaches, based on carbon nanotubes (1, 2), graphene (3-6), and dielectric-semiconductor architectures (7, 8), are promising as they show efficiencies of around 30 percent (7). Here, we achieve similar currents and voltages at modest flow velocities by replacing multistep fabrication (*e.g.* exfoliation, atomic layer deposition, chemical vapor deposition) with a single-step synthesis that is readily scaled to arbitrarily large areas. Nanolayers prepared from Fe and Ni produce several microA cm$^{-2}$, while Al and Cr nanolayers produce ten times smaller current densities, pointing to electron transfer within the thermal oxide nano-overlayers terminating the metals as an operational requirement that augments mere contact electrification on the devices by over an order of magnitude.

We prepare high-purity, all-inorganic single-element nanolayers by physical vapor deposition (PVD) of a standard-purity inexpensive metal (Fe, Ni, V, Al, Cr) onto solid or flexible substrates (glass, plastics, polymers) and then passivating the metal nanolayer in ambient air to form its own thermal oxide nano-overlayer. As is shown below, current is generated when moving aqueous droplets across the nanolayers. Moreover, use of a liquid flow cell overcomes the performance limitations inherent to aqueous droplets due to drop size (surface tension of water) and drop speed (terminal velocity on earth) and produces electrical currents when flowing aqueous solutions of alternating salinity across the device, or when running the device at constant salinity but alternating flow direction.

In the experiments, we form single- and dual-element nanolayers from low-cost 99.95% purity iron, 99.98% Ni, 99.7% V, 99.9995% aluminum, and 99.994% chromium sources. X-ray photoelectron spectroscopy (XPS) reveals a lack of common low-boiling point contaminants like calcium, magnesium, sodium, or zinc in the iron nanolayers and shows the presence of an ~3 nm thin oxidized iron nano-overlayer (9). Grazing incidence angle X-Ray diffraction experiments indicate the presence of crystalline $Fe^0$ with low index faces exposed but no crystallinity of the iron oxide overlayer (9). Control experiments show that this nano-overlayer forms spontaneously when the iron nanolayer is exposed to air and remains stable over prolonged (years) periods of time. Raman and XPS spectroscopy of the iron nanolayers indicate that the oxide nano-overlayer is composed of some Fe (III), $Fe_3O_4$, and other forms of iron oxide (9). Given nm-scale spatial variation of the oxide nano-overlayer thickness revealed by atom probe tomography (APT) reconstruction (10) described in more detail below, we expect corresponding heterogeneities in the electrostatic potentials – and charge distributions – in the metal below as well (see simulation results below).

We prepared Fe:FeOx nanolayers having 5, 10, 20, and 50 nm thickness, which differ in their transparency   (Fig. 1*A*), as well as 5 and 20 nm thin Al:AlOx and 10 nm Cr:CrOx, V:VOx, and Ni:NiOx nanolayers. Nanolayers were deposited onto 3 x 1 in$^2$ as well as 3 x 9 inch$^2$ glass microscope slides. The small slides were placed into a small Teflon cell containing a flow channel  (6 mm x 7.5 mm x 35 mm) Viton-sealed to the metal nanolayers  (Fig. 1*B*). The large slides were covered with a 1 mm thick silicone sheet into which a 180 mm x 15.2 mm wide opening was cut  (*SI Appendix*, Fig. S1) that was then covered by a 1 x 3 x 8 in$^3$ Kalrez block containing an in- and outlet fitting (NPT) to connect to a dual pump flow system and waste. After layering a second silicone sheet and a plexiglass cover on top, this large cell was sealed using large-area mechanical clamps.

Aqueous solutions consisted of deionized water containing varying amounts of NaCl, equilibrated with ambient air, thus reaching a pH of 5.8. For higher salt concentrations up to 2 M, we adjusted the pH to 8, given the relevance to ocean water and brine. "Instant Ocean" was used as well. Second harmonic generation $\chi^{(3)}$ measurements (9, 11) of the iron nanolayer indicate a negative interfacial charge density of -0.007 (3) C m$^{-2}$ at pH7, consistent with a considerable number density of deprotonated Fe-OH groups at the oxide/water interface near neutral pH. The change in interfacial electrostatic potential, $\Phi(0)$, or "gate" voltage, estimated from Gouy-Chapman theory, would then be in the -100 mV range when changing the salt concentration from 0.1 mM to 1 M.

When flowing water of alternating salinity at 20 mL min$^{-1}$ across a ~10 nm thin Fe:FeOx nanolayer in the small cell, currents of ~0.2 μA  (Fig. 2*A*) and voltages in the mV range are recorded. Currents approaching 1 µA are obtained in the large cell (Fig. 2*B*, note that the ionic strength gradient in the large cell is about ten times larger than that of the small cell, *vide infra*). When periodically alternating the direction of aqueous flow at constant ionic strength and constant flow rate in the large cell, current is generated as well (Fig. 2*C*), albeit in an asymmetric I vs t pattern attributed to the differences in inlet vs outlet size in the flow cell used. Current is also generated when alternating aqueous solutions of 600 mM NaCl with air (*SI Appendix*, Fig. S2), albeit at a smaller magnitude compared to continuous aqueous flow.

    Controls using open circuit voltage measurements show that increasing the iron layer thickness to 50 nm (from ellipsometry) leads to considerably smaller open circuit potentials when compared to thinner layers, while commercially available aluminum foil, aluminized polypropylene constituting a snack bag wrapping (~100 nm metal layer), a 2 mm thick sheet of iron metal (Alfar Aesar, 99.995%) or aluminum containing its native (thermal) oxide layer show no induced voltage (*SI Appendix*, Fig. S3). When using drops as opposed to continuous aqueous flow, we found that measuring the potential across as opposed to along the direction of drop motion shows little voltage during drop motion, and that reversing the polarity of the probes reverses the sign of the measured open circuit potential (*SI Appendix*, Fig. S4). 0.6 M Salt solutions representing the salinity of ocean water induce larger voltages than 0.1 M salt solutions that are comparable to those when using "Instant Ocean" (*SI Appendix*, Fig. S5). Alternating the drop salinity between that of the ocean (0.6 M, pH 8) and rainwater (0.2 mM, pH 5.8) induces regular current spikes over >8 hours (*SI Appendix*, Fig. S6). Using the small flow cell, the dynamics of the current flow can be correlated with the flow dynamics

inside the flow cell (*SI Appendix*, Fig. S7) for further improvement. Still frames from video recordings using clear and purple-colored water sources reveal a sharp concentration gradient in the flow cell during the time of maximum current generation (*SI Appendix*, Fig. S7), from which the "gate" footprint is conservatively estimated to be 7.5 mm channel width x 2 mm gradient width for subsequent estimations of current density, *j*, in the small cell. A similar analysis of the gradient in the big cell shows its footprint is ~2 $cm^2$. Alternating the salinity in drop experiments (*SI Appendix*, Fig. S7) produces several tens of mV in open circuit potential that are stable for hours. Additional experiments show induced currents and voltages with an external load resistance of up to 0.5 megaohm placed in series with the nanolayer. Of over 100 metal nanofilms prepared for this study, each produced comparable current (within a factor of 2) for comparable conditions of nanolayer thickness, flow cell dimensions, flow velocity, aqueous phase composition, and metal type.

    To gain a mechanistic understanding of current generation in the metal nanolayers, we carried out a series of experiments, as described next. Fig. 3*A* shows that Fe:FeOx, Ni:NiOx, and V:VOx nanolayers of 10 nm thickness produce currents that increase linearly with increasing flow rate at a rate of ~1 to ~3 microA $cm^{-2}$ per $cm \cdot s^{-1}$ increase in flow rate. The induced current densities are comparable to what can be achieved with falling water drops (blue vertical line). The produced currents are also comparable to what has been reported previously (3, 6) but obtained with considerably lower flow velocities when using 10 nm or 30 nm thin iron nanolayers or 10 nm thin nickel or vanadium nanolayers (*SI Appendix*, Fig. S10). Given that the iron oxide nano-overlayers contain iron in multiple oxidation states, we proceeded to investigate whether

metal nanolayers terminated with redox-inactive oxides would produce smaller currents. Indeed, Fig. 3*A* shows that 10 nm thin metal nanolayers prepared from Cr and Al produce considerably less current than 10 nm thin nanolayers prepared from Fe, Ni, or V at comparable flow conditions Fig. S8 shows a 20 nm Al:AlOx nanolayer also produces considerably less open circuit potential than the Fe:FeOx, Ni:NiOx, or V:VOx layers of comparable thickness. These results are rationalized by the observation that the iron, vanadium, and nickel nanolayers are terminated by thermal oxides that contain Fe(II) and Fe(III), V(IV) and V(V), and Ni(II) and Ni(III), respectively, whereas the aluminum and chromium metal nanolayers are terminated by thermal oxides that only contain metal in the +3 oxidation state (*SI Appendix*, Fig. S11).

As perhaps expected, the absence of metal results in negligible current, as shown in Fig. 3*B* for a 10 nm thin nanolayer of FeOx (no Fe(0) present) prepared by high-temperature quantitative ozone oxidation of a 10 nm thin nanolayer of Fe:FeOx. Fig. 3*C* shows that a 10 nm thin Fe:FeOx structure produces the highest currents when compared to thinner (5 nm) or thicker (30 nm and 50 nm) layers.

Given the results with the six different systems described in Fig. 3*A-B*, we expect that covering an active nanolayer (Fe:FeOx or Ni:NiOx) with a less active one (Al:AlOx or Cr:CrOx) should diminish the current density. Indeed, coating a 30 Fe:FeOx nanolayer with 5 nm Cr:CrOx results in considerable current reduction when compared to the neat Fe:FeOx nanolayer (Fig. 3*D*).

Taken together, the data shown in Fig. 3*A-D* provide evidence that intra-oxide electron transfer between $M^{m+}$ and $M^{n+}$ contributes to the current generation to a larger extent than would be expected from image charge formation alone in metal layers terminated by a redox inactive thermal oxide. Moreover, we expect that current generation can be further optimized by varying the nature and thickness of the metal and metal oxide layers in mixed metal architectures, alloys, or patterned nanolayers.

Our experiments additionally support the notion that surface charging of the metal oxide surface is an important part of the current generating mechanism in the metal nanolayers reported here. To explore this hypothesis, we recorded the electrical current as a function of the change in surface potential that occurs when changing the ionic strength from low to high salt concentration. To do so, we measured the current while changing the ionic strength from a given low salt concentration, say 0.1 mM, to 1 mM salt for several cycles, and then repeated those measurements for increasingly higher salt concentrations, each time starting at 0.1 mM (Fig. 3*C*). The largest currents are induced when the ionic strength difference is largest for each system studied. We then used experimental surface charge density estimates from second harmonic generation $\chi^{(3)}$ measurements (9, 11) to compute the change in Gouy-Chapman surface potential at the oxide/water interface for each ionic strength difference. Fig. 3*D* shows that the slopes in these "Tafel" plots are 110 (+/-20) $V^{-1}$ for the Fe:FeOx system. The Al:AlOx system, which is redox inactive under the conditions of our experiments, shows a slope of 7 (+/-2) $V^{-1}$ for all Gouy-Chapman surface potential differences surveyed except the highest, underscoring the large differences between the surface charging of the Al:AlOx and Fe:FeOx nanolayers.

Zooming out, Fig. 4 offers the following phenomenological interpretation of our findings, followed by a detailed microscopic investigation below. At the pH values used here (5.8 for low and 8.0 for high-salinity water), the water:oxide interfaces we investigate are charged (9, 12-17). The electrostatic potential reaches not only into the aqueous solution but also into the oxide (18, 19), depending on the local dielectric properties (20). Thus, if the oxide nano-overlayer is thin enough, the electrostatic potential extends beyond it to polarize the underlying metal, similar to metal atom charging on ultrathin oxides by underlying metals (21) or the phenomenology of the Cabrera-Mott model (22). Given the presence of different oxidation states in the iron (9), vanadium, and nickel oxide nano-overlayers, conduction by intra-oxide electron transfer, like what is known from bulk hematite crystals (23) or from chemical reactions on nanolayer metal-semiconductor heterostructures (24), is likely to be important as well. Electrical current is then generated by moving an electrical double layer (EDL) gradient (a "gate") across the metal:metal oxide nanolayer to drive electron transfer within the oxide nano-overlayer, which is coupled to the underlying metal nanolayer. The sharper the gradient, the larger the current density, $j$. Dendritic iron oxide features of ~5 nm x ~10 nm size (Fig. 5*A*) that extend from the surface into the bulk of the iron metal nanolayer, as revealed by APT (10), may open possibilities for an electron and/or hole ratchet, similar to what has been proposed for low-light energy-driven transducers (25), or pose limits due to tunneling losses. Structures whose oxide nano-overlayers contain only a single oxidation state, such as those formed from Al or Cr metal, should still produce currents due to contact electrification, but the lack of intra-oxide electron transfer would diminish their current output.

The system presented here differs in several aspects from recent demonstrations of flow-induced power generation. First, our experiments are consistent with a mechanism for electrical current generation that involves redox activity in the metal oxide layer, which is reported here for the first time. Second, the all-inorganic devices described here consist of metal nanolayers formed on a given support in a single step over arbitrarily large areas using an electron beam deposition apparatus. Upon exposure to ambient air, an oxide nano-overlayer forms spontaneously and then self-terminates after ~3 to ~5 nm, depending on the thickness of the underlying metal nanolayer. The high purity of the metal nanolayer prevents further growth of the oxide nano-overlayer, resulting in a stable structure. Third, the amphoterism of the thermal oxide nano-overlayer is critical to EDL gradient, or "gate", formation as solutions move across the liquid:solid interface. Fourth, the thickness of the metal nanolayer needed to produce current (Fig. 3*C*) is comparable to the mean free path of the electrons in it (26), engendering a propensity for charge motion parallel to as opposed to away from the interface. Fifth, the starting materials, a suitable support and a standard-purity metal source (Fe, Ni, V, Al, Cr etc.), are inexpensive, costing ~$1 to $2 g$^{-1}$.

To probe the charge fluctuations in the metal:metal oxide (M:MOx) nanolayer in the presence of moving ions, calculations were performed using an all-atom molecular dynamics (MD, see *SI Apendix* Note 2 and *SI Appendix* Table S1) model for the solvent, ions, and a M:MOx nanolayer, including charge-polarization of the nanolayer and image-charge interactions between the nanolayer and the solution. The M:MOx nanolayer is modeled after the APT reconstruction of the Fe:FeOx nanolayer (Fig. 5*A*) as a polarizable metal conductor (Fig. 5*B*, grey) with a non-polarizable oxide heterostructure (pink). The subsurface metal/oxide heterostructure is modeled in a simple columnar geometry with a range of values for the width, *d*. For a given width of the oxide heterostructure (*d*=1.3 nm), Fig. 5*C* illustrates the distribution of induced charge in the nanolayer for several positions of a sodium cation. Substantial polarization of the metal for ion positions away from the nonpolarizable heterostructure is reduced when the cation is positioned above the heterostructure (Fig. 5*C* and *SI Appendix*, Fig. S12). This position-dependence of the induced charge manifests in the Coulomb interaction between the ion and the nanolayer (Fig. 5*D*, leading to a heterostructure-dependent interaction potential between the M:MOx nanolayer and the ion, with a potential energy barrier appearing in the region of the nonpolarizable heterostructure.

To examine these nanolayer polarization effects in the presence of a solution with alternating salinity, Fig. 5*E* shows a snapshot of all-atom MD simulations, with vertical lines indicating semipermeable boundaries for the solvated ions and with the instantaneous induced charge fluctuations on the electrode shown in red-blue scale. Fig. 5*F* shows the time-averaged (black) charge induced charge distribution for the shown simulation cell, as well as 0.5 ns block-averages of the distribution (other colors). Two features are immediately clear: *(i)* the induced charge distributions in the metal/oxide nanolayer undergo dramatic fluctuations with changes of the ion and water configuration, which reflect changes in the transient electrostatic interactions between the nanolayer and the solvated ions, and *(ii)* these induced charges are massively damped out in the vicinity of the nonpolarizable heterostructure, i.e. the oxide nano-overlayer. Fig. 5*G* shows that the effect of the heterostructure on the average induced charge is much smaller than its effect on the fluctuations.

The simulations in Fig. 5*F-G* reveal that the nonpolarizable heterostructure model of the metal oxide nano-overlayer creates spatial variation in the local induced charge fluctuations in the metal nanolayer below. These fluctuations are proportional to the local interfacial capacitance, i.e., $C_F = \beta \langle (\delta Q(x))^2 \rangle$ (27). Given that this interfacial capacitance connects droplet motion to induced current, $I = -\psi \frac{dC_F}{dt}$ where $\psi$ is the surface potential (see equivalent circuit in *SI Appendix* Note 3 and Fig. S13 and also in Ref. (3)), the simulations thus provide a direct connection between the morphology of the oxide heterostructure and the gate-induced current presented here. Moreover, these simulations reveal that the interfacial capacitance that gives rise to the current is strikingly sensitive to the electronic character and spatial features of oxide heterostructure, such that nanometer-scale changes in the heterostructure give rise to unexpectedly large effects in the resulting interfacial capacitance.

The effects observed in the simulations are expected to be further enriched by the amphoterism of the oxide overlayer, which is important for determining the sign and magnitude of the charge and potential distributions within the EDL under conditions of varying aqueous pH and ionic strength. Control over the structure of the oxide dendrites, their number density, and their width and depth offers the possibility to further optimize charge mobility along the potential hotspots on the dendrites and minimize possible leakage due to tunneling. Additional control comes from the choice and concentration of ions in the aqueous phase and the steepness of the salt concentration gradient, which determines the area of the gate footprint at the aqueous/solid interface (steeper gradients lead to increased current densities, *j*). Moreover, the volcano plot-like current vs M:MOx film thickness data shown in Fig. 3*C* suggests that film thickness on the order of the mean free path of the electron are required for current generation, offering an additional means of optimizing the electron current flow.

The relatively modest flow velocities surveyed here (a few cm s$^{-1}$) indicate the approach presented here may work in entirely passively operating assemblies, yet, there is ample room for improvement. The use of appropriate metals having biocidic properties (Al, Zn, Ag, Cu) may have the additional benefit of counteracting biofilm formation in the field. The optical properties of the iron and nickel nanolayers also open the possibility of further charge carrier generation by visible light, conversion boosting of solar cells, or the coating of building windows with them, given the ionic strength of rainwater (0.2 mM) (13) and the low absorbance of the nanolayers in the visible. PVD onto plastics or flexible substrates (*SI Appendix* Note 4 and Fig. S14 and S15) also allows for large-area yet light-weight and/or foldable designs. PVD of appropriately formulated metal nanolayers into tubes allows for implantable applications *in vivo*, while PVD of metal nanolayers onto a range of other polymers we surveyed opens the door to transducers operating in three-dimensional structures prepared, for instance, by 3D printing. Despite these exciting possibilities, we caution that the underlying mechanism needs to be further investigated, for instance, by specifically following the charge carrier motion in real time and space.

**Methods.** The nm-thin iron layers and their oxide nano-overlayers were prepared on glass microscope slides (VWR) and characterized as described in our previous work (9, 28). We used computer controlled multi-channel Ismatec peristaltic pumps (ISM4408). Aqueous solutions were prepared from NaCl (Sigma-Aldrich) in Millipore water adjusted to pH 7, or equilibrated with ambient air to pH 5.8, and containing various amounts of NaCl, as indicated in the relevant Figure Captions. "Instant Ocean Aquarium Sea Salt" was used as received from Amazon (ASIN: B00NQH210G). The drop experiments were performed using motorized syringe pumps (Harvard Apparatus Elite 11). Using Teflon tubing, drops having an average volume of 0.0165 (1) mL (measured for a flow rate of 0.5 mL/min) were released in ambient laboratory air from a height of 10 cm onto a given device held in air by an electrically insulated clamp at an incident angles of ~20 degrees. Variations in incident angle, drop release height, and drop size lead to variations in drop flow dynamics and velocity on the nanolayer surfaces and corresponding variations in magnitude and duration of the measured open circuit voltage spikes, similar to what had been reported in the earlier studies using carbon- and semiconductor structures that are mentioned in the Main Text. Nanolayers stored for prolonged periods of time (~two years) in ambient laboratory air showed larger contact angles (Computerized First Ten Ångstroms contact angle goniometer, $\theta$=57±5° from seven replicates using deionized water) than freshly prepared nanolayers ($\theta$=37±3° from seven replicates using deionized water), on which the water drops spread considerably more while also producing open circuit potential spikes that are somewhat larger in magnitude and longer in duration (*SI Appendix*, Fig. S9). Given the potential relevance of our system for use in the ambient environment, we emphasize here the results from nanolayers that had been stored in the dark for about two years in conical centrifuge tubes made of polypropylene (Falcon, 50 mL, with screw top) containing ambient air. Drops rolling off the device were collected in a receptacle. Open circuit potential measurements were performed using a Keithley 2100 voltmeter and standard alligator clip-on probes, taking special care to keep the probes dry. The resistance of the dry nanolayers is around 50 to 500 Ohm (Keithley 2100), depending on layer thickness. Short circuit current measurements were carried out on an Agilent B1500A semiconductor parameter analyzer equipped with a high-resolution SMU, and on a Keithley 6485 Ammeter.

Details regarding the computational methods and the model we used are described in *SI Appendix* Note 2.

**Acknowledgments.** M.D.B. gratefully acknowledges support from the PPG fellowship program at Northwestern University. This work was supported by the US National Science Foundation (NSF) under its graduate fellowship research program (GRFP) award to P.E.O. We also acknowledge support from Northwestern University's Presidential Fellowship program (P.E.O.), the Center for Water Research (E.J.L.), the Undergraduate Research program (C.E.W.), and the Dow Professorship program (F.M.G.). We are thankful to Dr. Wei Huang for the assistance with the first current measurements on the Agilent B1500A. F.M.G. gratefully acknowledges support from the NSF through award number CHE-1464916 and a Friedrich Wilhelm Bessel Prize from the Alexander von Humboldt Foundation. T.F.M. acknowledges support from the Office of Naval Research under Award No. N00014-10-1-0884. F.M.G. and T.F.M. acknowledge support from DARPA through the Army Research Office Chemical Sciences Division under Award No. W911NF1910361.

**Author Contributions.** F.M.G. conceived of the idea. M.D.B., E.H.L., P.E.O., C.E.W., and F.M.G. executed the experiments and analyzed the data. J.K. and T.F.M. performed the computer simulations and analyzed the data. The manuscript was written with substantial contributions from all authors.

**Supplementary information available.** Computational details and results from the controls mentioned in the main text.

**Author Information.** The authors declare no competing financial interests. Correspondence should be addressed to FMG (geigerf@chem.northwestern.edu).

Boamah *et al*.                                                                                                              Page 17**Data availability.** All relevant data are available from the authors upon request to the corresponding author.

**References.**

1. S. Ghosh, A. K. Sood, N. Kumar, Carbon nanotube flow sensors. *Science* **299**, 1042-1044 (2003).
2. Z. Zhang *et al.*, Emerging hydrovoltaic technology. *Nature Nanotechnology* **13** (2018).
3. J. Yin, X. Li, Z. Zhang, J. Zhou, W. Guo, Generating electricity by moving a droplet of ionic liquid along graphene. *Nature Nanotechnology* **9**, 378-383 (2014).
4. S. Yang *et al.*, Mechanism of Electric Power Generation from Ionic Droplet Motion on Polymer Supported Graphene. *J. Am. Chem. Soc.* **140** (2018).
5. Q. Tang, P. Yang, The era of water-enabled electricity generation from graphene. *J. Mat. Chem. A* **4**, 9730-9738 (2016).
6. J. P. G. Tarelho *et al.*, Graphene-based materials and structuresfor energy harvesting withfluids–A review. *Mat. Today* **21** (2018).
7. J. Park *et al.*, Identification of Droplet-Flow-Induced Electric Energy on Electrolyte−Insulator−Semiconductor Structure. *J. Am. Chem. Soc.* **139**, 10968-10971 (2017).
8. X. Li *et al.*, Hydroelectric generator from transparent flexible zinc oxide nanofilms. *Nano Energy* **32** (2017).
9. D. Faurie-Wisniewski, F. M. Geiger, Synthesis and Characterization of Chemically Pure Nanometer-Thin Zero-Valent Iron Films and Their Surfaces. *The Journal of Physical Chemistry C* **118**, 23256-23263 (2014).
10. M. D. Boamah, D. Isheim, F. M. Geiger, Dendritic Oxide Growth in Zerovalent Iron Nanofilms Revealed by Atom Probe Tomography. *J. Phys. Chem. C* **122**, 28225-28232 (2018).
11. P. E. Ohno, S. A. Saslow, H.-f. Wang, F. M. Geiger, K. B. Eisenthal, Phase-referenced Nonlinear Spectroscopy of the alpha-Quartz/Water Interface. *Nature communications* **7**, 13587 (2016).
12. A. Adamson, *Physical Chemistry of Surfaces* (John Wiley & Sons:, New York, ed. 5th ed, 1990).
13. D. Langmuir, *Aqueous Environmental Geochemistry* (Prentice-Hall, Inc, New Jersey, 1997).
14. J. Lyklema, *Fundamentals of Interface and Colloid Science* (Elsevier, 2000).
15. G. E. Brown, How Minerals React with Water. *Science* **294**, 67-70 (2001).
16. C. Macias-Romero, I. Nahalka, H. I. Okur, S. Roke, Optical imaging of surface chemistry and dynamics in confinement. *Science* **25**, 784-788 (2017).
17. H. J. Butt, K. Graf, M. Kappl, *Physics and Chemistry of Interfaces* (Wiley VCH, Weinheim, 2013).
18. S. M. Sze, K. K. Ng, *Physics of Semiconductor Devices* (John Wiley & Sons, Hoboken, NJ, ed. 3rd, 2007).
19. X. Yu, T. J. Marks, A. Facchetti, Metal oxides for optoelectronic applications. *Nature Mater.* **15** (2016).
20. W. M. Telford, L. P. Geldart, R. E. Sheriff, *Applied Geophysics, Chapter 5: Electrical Properties of Rocks and Minerals* (Cambridge University Press, 1990).
21. G. Pacchioni, L. Giordano, M. Baistrocchi, Charging of Metal Atoms on Ultrathin MgO/Mo(100) Films. *Phys. Rev. Lett.* **94**, 226104 (2005).

Figure Legends

**Fig. 1: Metal nanolayers for gravitational to electrical energy conversion.** (*A*) Photographs iron and aluminum nanolayers with indicated thicknesses on microscope glass slides over the Northwestern University seal. (*B*) Photograph of Teflon cell with flow channel. Dashed lines indicate substrate position and arrows indicate aqueous flow direction.

**Fig. 2: Current and Voltage Measurements**. (*A*) Current induced in a 10 nm Fe:FeOx nanolayer (3x1 in$^2$) when flowing deionized (DI) water at pH 5.8 for 20 sec (blue segment), followed by 20 sec flow of 1 M NaCl held at pH 7 (green segment), and six subsequent replicates, all at a constant flow rate of 20 mL min$^{-1}$. (*B*) Same as in (*A*), but measured using a 3x9 in$^2$ Fe:FeOx nanolayer of 10 nm thickness at a flow rate of 100 mL min$^{-1}$ and 2 min between switching salt concentration. (*C*) Same as in (*B*), but measured at a flow rate of 35 mL min$^{-1}$ and constant 0.6 M salt concentration while reversing the flow direction every 2 min, marked by the vertical dashed lines.

**Fig. 3: Mechanistic Investigations.** (*A*) Average current densities measured as a function of aqueous flow velocity using 10 nm thin nanolayers of Fe:FeOx (blue-filled circles), Ni:NiOx (purple-filled circles), V:VOx (red-filled circles), Al:AlOx (grey-filled circles), and Cr:CrOx (orange-filled circles) while alternating deionized water (pH=5.8) and 0.6 M NaCl solution (pH ~7) segments every 20 sec, and current density obtained for 30 μL drops falling with a 0.1 to 0.2 cm$^2$ contact area onto a 10 nm thick Fe:FeOx nanolayer deposited onto a 1x3 in$^2$ glass substrate while alternating the drop salinity between deionized water and 0.6 M at a drop rate of 2 mL min$^{-1}$ and an incident angle of ~160° (vertical blue bar). Error bars on point estimates shown are for 1 standard

deviation (σ) from n=7 and 8 replicate measurements per flow rate. (*B*) Same a (*A*), but for a 10 nm Fe:FeOx nanolayer (blue-filled circle) as a function of aqueous flow velocity and for a 10 nm thin nanolayer of pure FeOx (orange-filled circle, no metal present) and a 10 nm thin nanolayer of pure TiOx (gray-filled circle). (*C*) Current density recorded for Fe:FeOx nanolayers varying in total thickness obtained with a flow velocity of 0.74 cm s$^{-1}$ while alternating deionized water and 0.6 M NaCl solution segments every 20 sec. (*D*) Current density obtained for a 30 nm Fe:FeOx nanolayer without (deep blue-filled circle) and with a 5 nm Cr:CrOx nanolayer on top of it (yellow-stroked deep blue-filled circle) obtained with a flow velocity of 1.15 cm s$^{-1}$, and for a 30 nm nanolayer of pure FeOx (no metal present) obtained with a flow velocity of 1 cm s$^{-1}$, all while alternating deionized water and 0.6 M NaCl solution segments every 20 sec. (*E*) Current density for Fe:FeOx (filled and empty blue circles for 10 and 5 nm thickness, respectively) and Al:AlOx (empty grey circles, 20 nm thickness) nanolayers as a function of natural logarithm of the salt concentration difference in solutions of alternating salinity recorded using 30 μL drops at a drop rate of 2 mL min$^{-1}$ (flow velocity=0.3 cm s$^{-1}$, assuming a 0.1 cm$^2$ contact area of the rolling drop). Error bars on point estimates shown are for 1 standard deviation (σ) from n=O (100) replicate measurements. (*F*) Natural logarithm of the current density (in A cm$^{-2}$) as a function of change in Gouy-Chapman surface potential (σ=0.007 C m$^{-2}$) resulting from changing the ionic strength when altering the salt concentration.

**Fig. 4:** Cartoon representation of electrical energy conversion in metal nanolayers terminated by their thermal oxides.

**Fig. 5: Model of Charge Mobility in Nanoconfined, Insulator-Terminated Metal Conductor.** (*A*) Atom probe tomography reconstruction of the heterostructured Fe:FeOx nanolayer (center). Iron oxide and iron metal shown separately on top (red) and bottom (blue), respectively. (*B*) All-atom representation of the heterostructured nanolayer, including the metal conductor (gray) and a nonpolarizable oxide overlayer and with columnar subsurface heterostructure (pink); a single probe Na+ cation is shown at a distance of 1.6 angstroms from the nanolayer. (*C*) Induced charge distribution, Q (x), by the Na$^+$ cation at four different lateral positions relative to the position of the nonpolarizable heterostructure. (*D*) Ion-nanolayer Coulomb interaction as a function of function of lateral ion position, for various widths, *d*, of the nonpolarizable heterostructure; $\Delta E^{coul}$ is the difference in the ion-nanolayer Coulomb interaction for the nanolayer systems with and without the subsurface heterostructure. (*E*) MD simulation snapshot for alternating regions of ionized (0.43 M NaCl) water / de-ionized water in contact with the nanolayer with columnar heterostructure (*d*=1.3 nm). The nanolayer is shown as in (*B*), but with the in stantaneous charge polarization of metal conductor atoms also indicated (range = [-0.005 *e* (blue), +0.005 *e* (red)]). Vertical dotted lines indicate semipermeable boundaries for the ions to preserve the salinity boundaries. (*F*) For the simulation cell shown in (*E*), the time-averaged induced charge distribution, Q (x) (black), as well as the 0.5-ns block averages of the same quantity. (*G*) Comparison of the time-averaged induced charge distribution for the system with (black) and without (red) nonpolarizable heterostructure.

A

5 nm Fe

10 nm Fe

20 nm Fe

20 nm Al

50 nm Fe

B

Flow out

4 cm

Flow in

1 cm

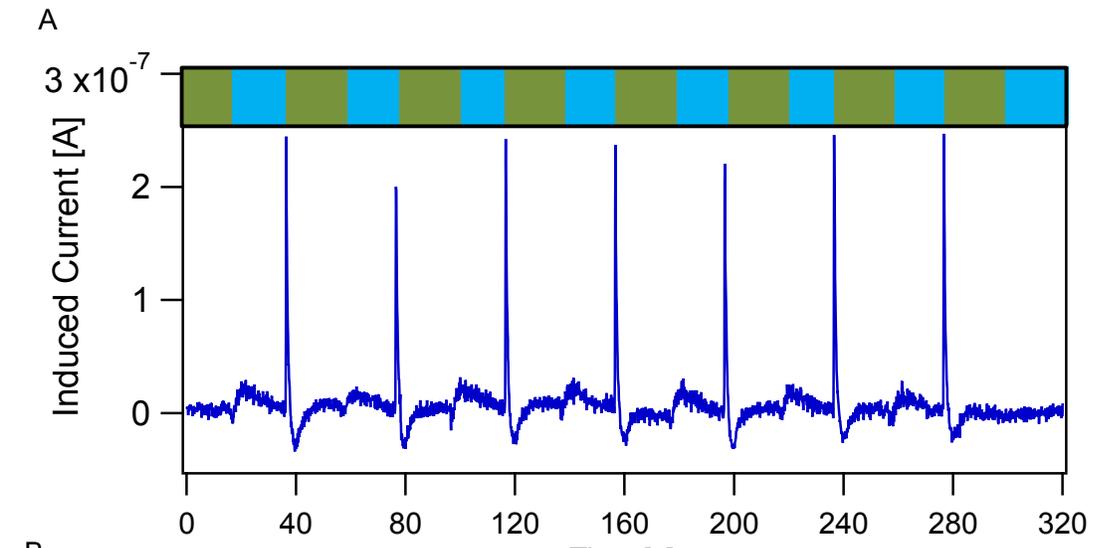

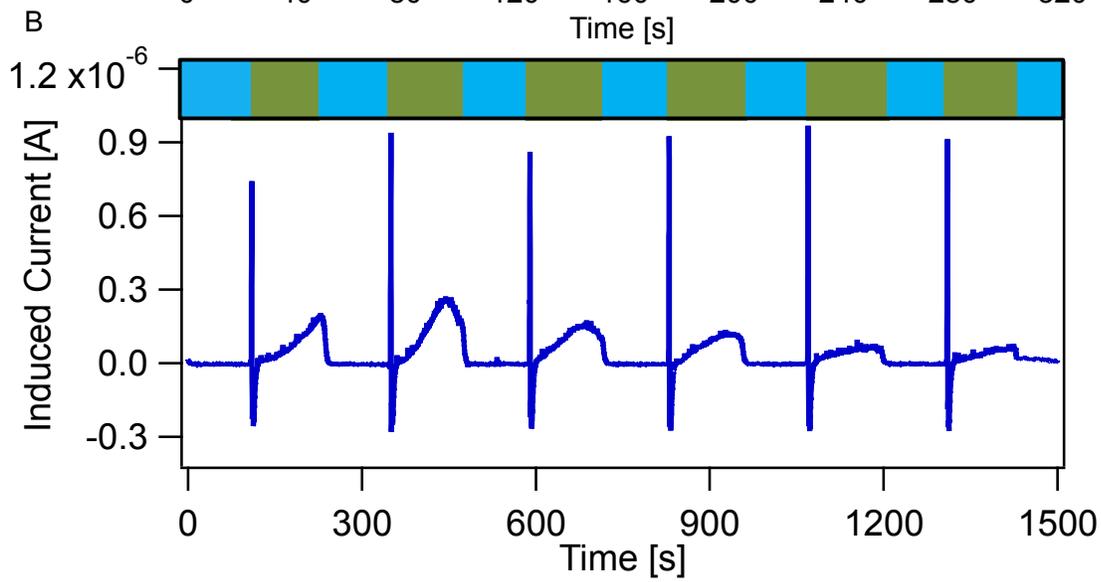

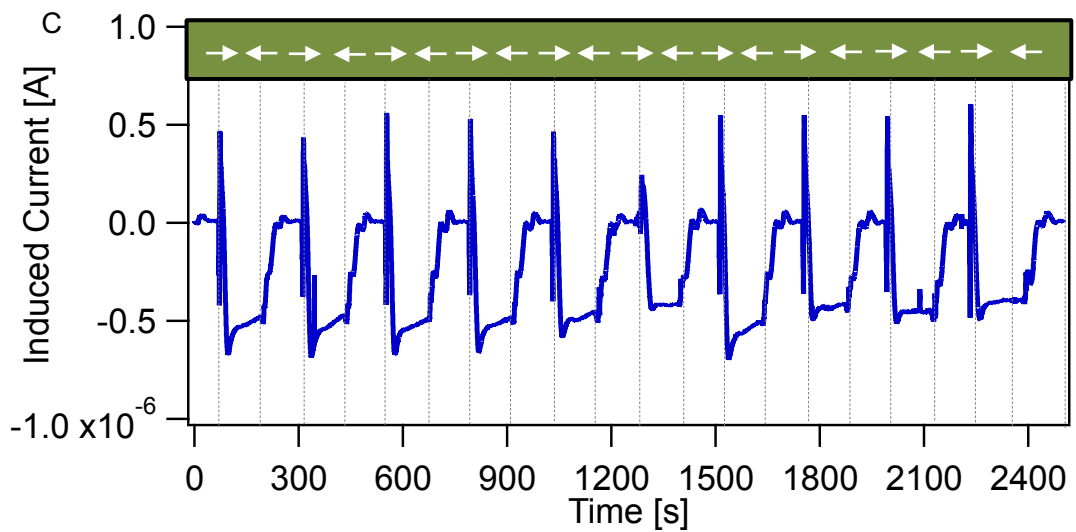

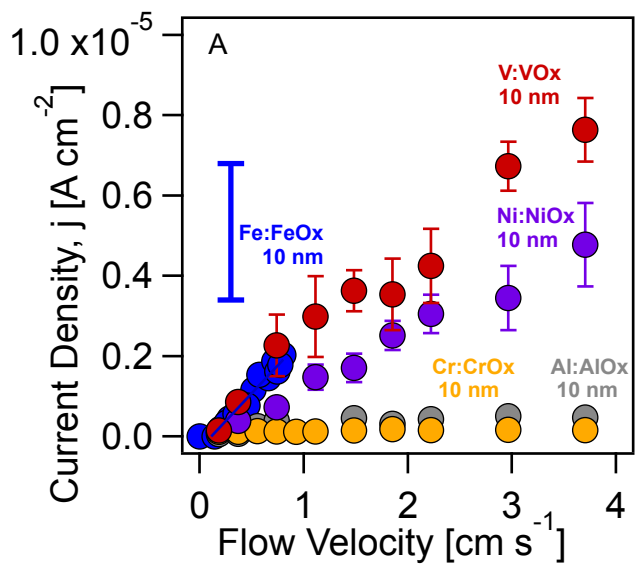
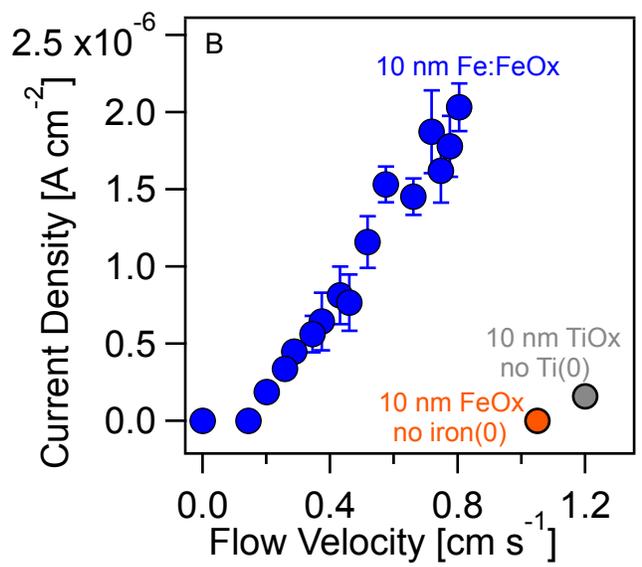
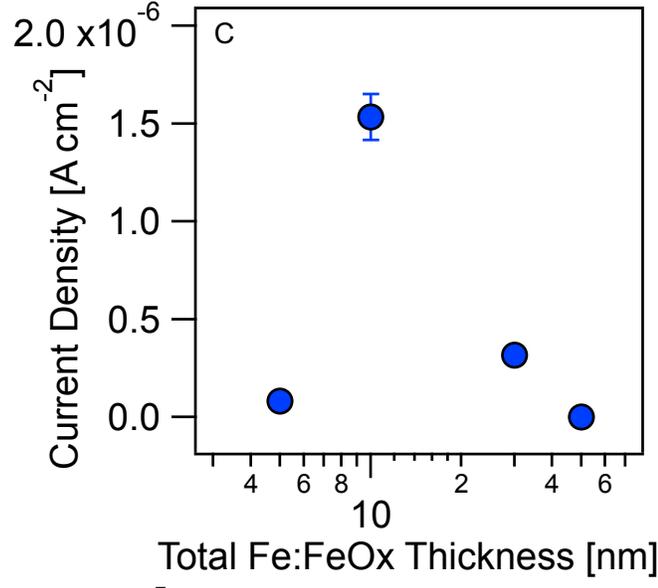
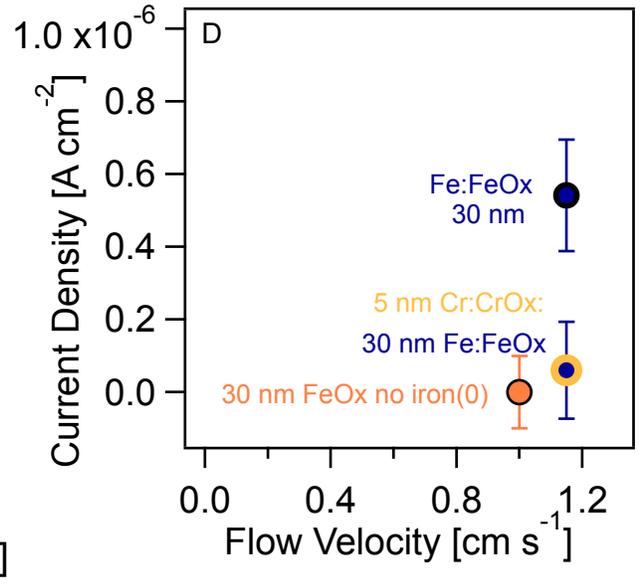
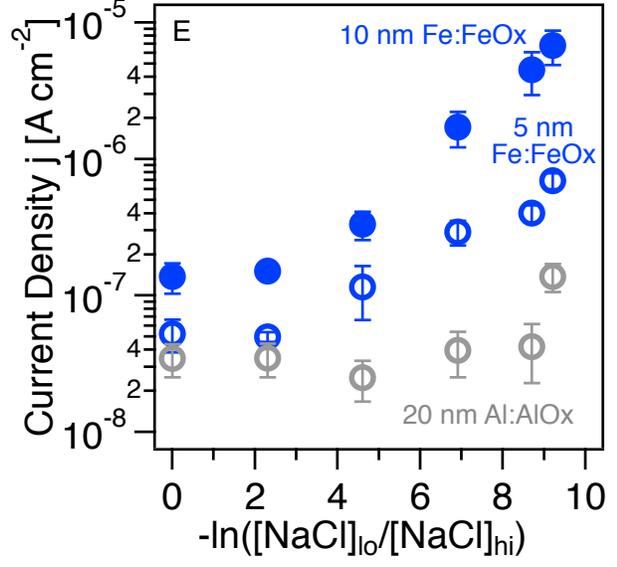
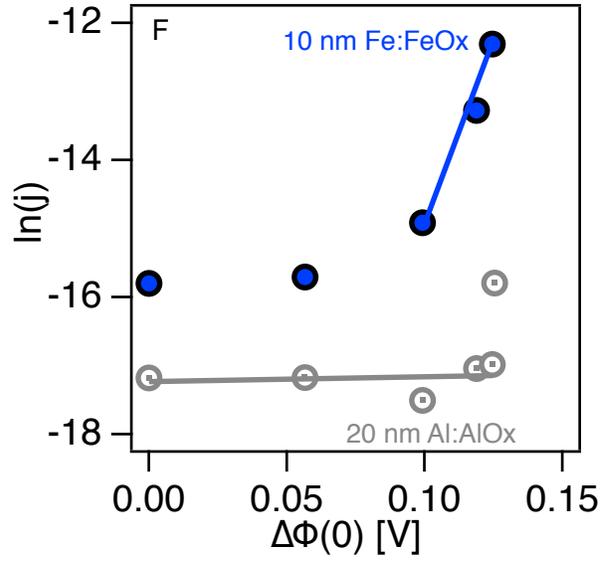

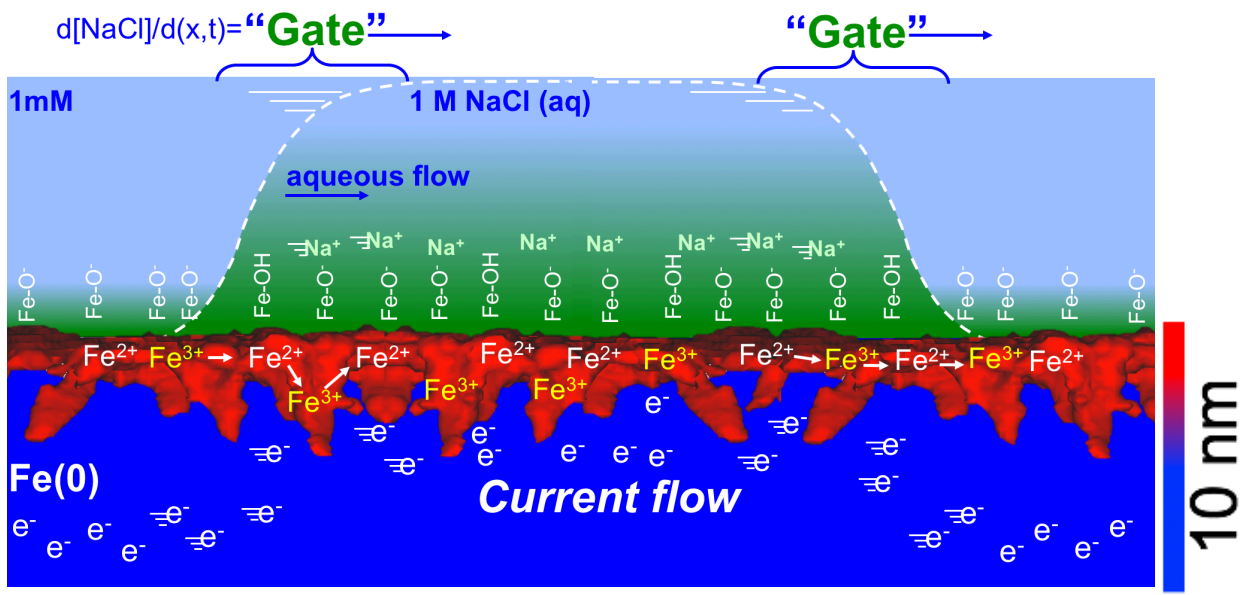

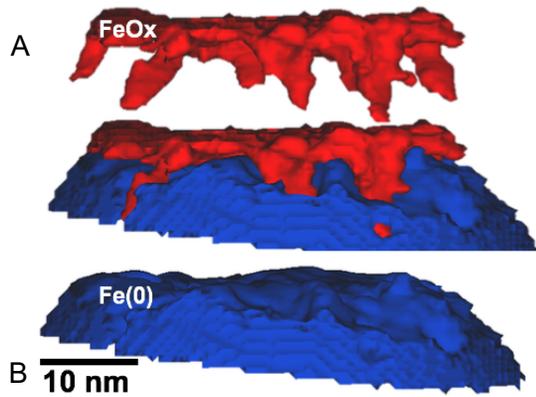

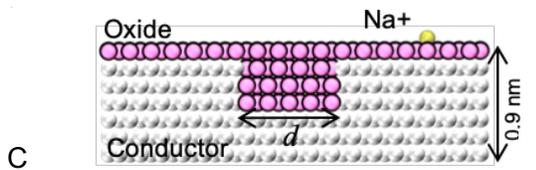

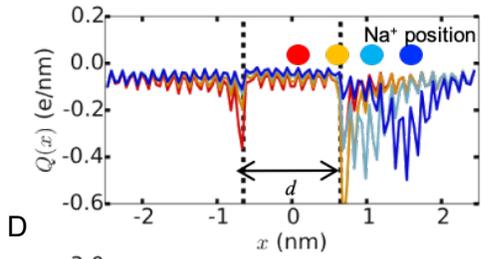

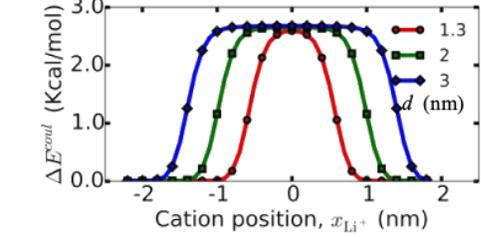

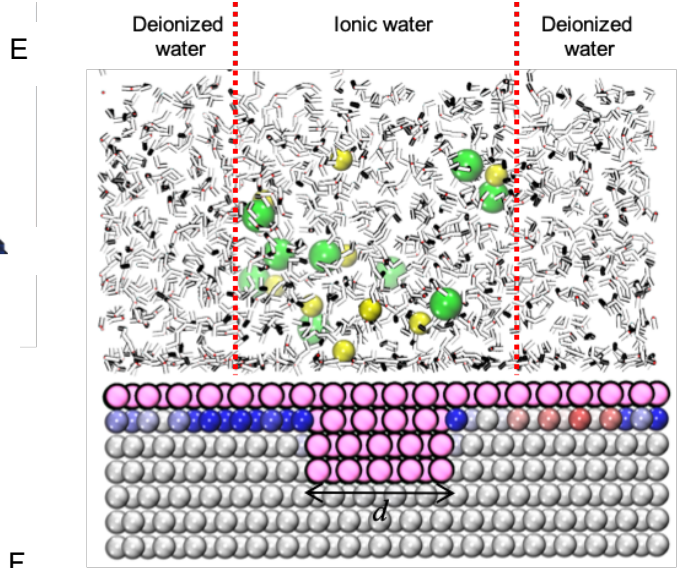

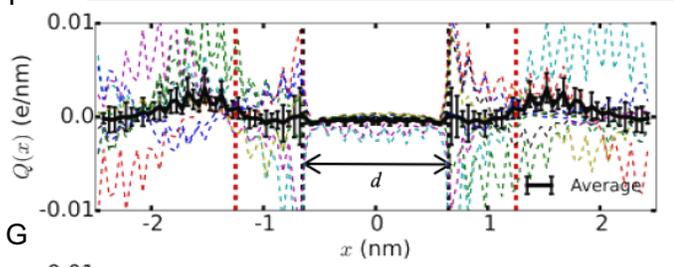

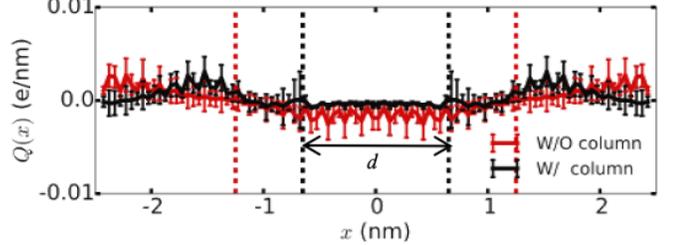